\documentclass[aps,pre,twocolumn,showpacs]{revtex4}
\usepackage{graphicx,bm,amsmath,dcolumn,}
\begin{document}
\newcommand{\be}{\begin{equation}}
\newcommand{\ee}{\end{equation}}
\newcommand{\half}{\frac{1}{2}}
\newcommand{\ith}{^{(i)}}
\newcommand{\im}{^{(i-1)}}
\newcommand{\gae}
{\,\hbox{\lower0.5ex\hbox{$\sim$}\llap{\raise0.5ex\hbox{$>$}}}\,}
\newcommand{\lae}
{\,\hbox{\lower0.5ex\hbox{$\sim$}\llap{\raise0.5ex\hbox{$<$}}}\,}
\newcommand{\mat}[1]{{\bf #1}}

\title{Crossover phenomena involving the dense O($n$) phase}
\author{ Wenan Guo$^{1}$ and Henk W. J. Bl\"ote$^{2}$ } 
\affiliation{$^{1}$Physics Department, Beijing Normal University,
Beijing 100875, P. R. China}
\affiliation{$^{2}$ Instituut Lorentz, Leiden University,
P.O. Box 9506, 2300 RA Leiden, The Netherlands}
\date{\today} 
\begin{abstract}
We explore the properties of the low-temperature phase of the O($n$)
loop model in two dimensions by means of transfer-matrix calculations
and finite-size scaling. We determine the stability of this phase with
respect to several kinds of perturbations, including cubic anisotropy,
attraction between loop segments, double bonds and crossing bonds. In
line with Coulomb gas predictions, cubic anisotropy and crossing bonds
are found to be relevant and introduce crossover to different types of
behavior. Whereas perturbations in the form of loop-loop attractions and
double bonds are irrelevant, sufficiently strong perturbations of these
types induce a phase transition of the Ising type, at least in the cases
investigated. This Ising transition leaves the underlying universal
low-temperature O($n$) behavior unaffected.
\end{abstract}
\maketitle 

\section{Introduction}
\label{intro}
The O($n$) spin model is defined in terms of $n$-component spins on a
lattice, with spin-spin interactions that satisfy O($n$) symmetry,
i.e., the model is isotropic in the $n$-dimensional spin-vector space.
The cases $n=1$, 2 and 3 correspond with the Ising, XY and Heisenberg
models respectively, but the significance of the O($n$) model goes beyond
these spin models.  A loop expansion of the partition function of certain
two-dimensional O($n$) spin models \cite{Stanley,Domea} leads to a system
of nonintersecting loops, while the spin degrees of freedom are integrated
out.  The resulting loop gas is called the O($n$) loop model, sometimes
abbreviated to just O($n$) model. It has only discrete degrees of
freedom, but the spin dimensionality $n$ appears in the partition sum
of the O($n$) loop model as a continuously variable parameter.
In the limit $n \to 0$, the model serves to describe the behavior of
polymer configurations \cite{deG,Np,DS}.

For some two-dimensional O($n$) spin models, a mapping on a loop model is
possible such that it yields the partition function in a form that enables the
derivation of exact results \cite{N,Baxter,BB,Suzuki,Kunz-Wu,BNW,3WBN,3WPSN}.
These results show that there exist several ``branches'' of universality
classes that continuously depend on the parameter $n$ for $-2\leq n\leq 2$.

One of these branches describes the phase transition between the
high-temperature disordered spin phase and the low-temperature phase,
where the spins display long-ranged correlations. In the terms of the
loop model, the high-temperature phase is characterized by small loops
and a low loop density, and the low-temperature phase by a high loop
density and the existence of a loop of divergent size.

The low-temperature phase appears to be more interesting than what one
might expect on the basis of the known properties of the long-range
ordered O(1) or Ising model. For general $n$ in the interval
$-2 \leq n \leq 2$, the low-temperature phase is still critical in the
sense that the correlation functions display power-law behavior. Its
universal properties are described by another exactly solved branch.
Recently, an exact transformation was applied to map the low-temperature
branch of the O($n$) loop model onto a tricritical loop model that includes
vacant sites \cite{NGB}. This mapping was applied for the case of the
honeycomb as well as for that of the  square lattice.
Since this tricritical O($n$)
model should have two more relevant temperature-like fields than the
low-temperature branch, one may wonder whether these relevant directions
have some physical meaning in the low-temperature O($n$) phase. 

The present work focuses on the stability properties of the
low-temperature phase of the loop model with respect to several
perturbations that move the loop model away from the exactly solvable
point.  These perturbations are:
\begin{enumerate}
\item
An attractive potential associated with loop segments that collide at a
vertex of the lattice;
\item
The introduction of double bonds, which allow some lattice edges to be
covered by up to two loop segments;
\item
A cubic perturbation of the O($n$) spin symmetry, which translates into
the connection of four incoming loop segments at a vertex;
\item
Crossing bonds coupling O($n$) spins, which correspond with crossing
loop segments in the loop model, without affecting the O($n$) symmetry
of the corresponding spin model.
\end{enumerate}
The existing results in the literature, in particular from Coulomb gas
theory \cite{Kad,CG} predict, or at least suggest, the effects of
these perturbations.
Cubic deviations from O($n$) symmetry were concluded to be irrelevant
on the critical branch for $n<2$, and to be relevant on the $n<2$
low-temperature branch \cite{CG}. Crossing bonds are predicted to be
described by the same exponent, so that they should also be
relevant in the low-temperature phase. Attractions between loop segments
were however concluded \cite{BN} to be irrelevant in this phase.

Our present work purports to test the theoretical predictions
numerically, by means of transfer-matrix calculations.
In Sec.~\ref{models} we define the models under investigation,
and summarize the relevant existing results. Section \ref{trmat}
explains the numerical procedures, and Sec.~\ref{numres} presents
the numerical results, concerning the phase diagram and the relevance
or irrelevance of the various perturbations.  We conclude with 
a discussion of the results in Sec.~\ref{disc}.

\section{Models}
\label{models}
The O($n$) spin model with pair interactions is described by the
reduced Hamiltonian
\be
{\mathcal H}=-\frac{1}{k_{\rm B}T} \sum_{<ij>}J(\vec{s}_i\cdot\vec{s}_j)\, ,
\label{Hon}
\ee
where the sum is over all nearest-neighbor pairs, and the $\vec{s}_i$ are
$n$-dimensional vectors whose label indicates the site number $i$. They
are normalized as $\vec{s}_i \cdot \vec{s}_i=n$, and their integration
measure is $\int d{\vec{s}_i}=1$. The function $J$ describes the
pair energy as a function of the spin product, and is usually chosen as
a multiplicative constant, although other choices still preserve the O($n$)
symmetry. For the special choice $J(y)= k_{\rm B}T\ln(1+zy)$ the Hamiltonian
becomes
\be
{\mathcal H}=-\sum_{<ij>} \ln  (1+z \vec{s}_i \cdot \vec{s}_j) \, ,
\label{Hons}
\ee
where the parameter $z$ represents the coupling strength between 
neighboring O($n$) spins, and can thus be understood as
a measure of the inverse temperature.
We consider the ferromagnetic case $z > 0$.
The partition integral of this model can be written as
\be
Z=  \left[\prod_{k} \int d{\vec{s}}_k \right]  \prod_{<ij>} (1+z
 \vec{s}_{i} \cdot \vec{s}_{i}) \, ,
\label{Pint}
\ee
where the products are on the sites and on the nearest-neighbor pairs of
the lattice respectively.
\subsection{Loop model on the honeycomb lattice}
For the model on the honeycomb lattice, a graph expansion \cite{Domea}
of Eq.~(\ref{Pint}) expresses the partition function in terms of a
sum over all configurations ${\mathcal G}$ of nonintersecting loops on
the edges of the honeycomb lattice:
\begin{equation}
Z_{{\rm loop}} =\sum_{{\mathcal G}} z^{N_b} \, n^{N_l} \, ,
\label{Zloop}
\end{equation}
where the graph ${\mathcal G}$ covers $N_b$ bonds of the lattice, and
consists of $N_l$ closed, nonintersecting loops. Each lattice edge may
be covered by at most one loop segment.
Exact analysis  \cite{N,Baxter,BB,Suzuki} appears possible for special
values $z = z_{\rm c}$  given by
\be
z_{\rm c} =1/\sqrt{2 \pm \sqrt{2-n}}\, , ~~~~~-2\leq n \leq2 \, ,
\label{hczc}
\ee
where the plus sign corresponds with a critical ordering transition
separating the high-temperature phase from the low-temperature phase.
The minus sign corresponds with the low-temperature O($n$) phase.
The solutions with the plus-sign are called branch 1, those with the
minus sign branch 2. The exact results include the leading scaling
dimensions.

These theoretical analyses are possible because of the special form of
the spin-spin interaction $J$ in Eq.~(\ref{Hon}) and because each spin
occurs at most to the third power in the expansion of Eq.~(\ref{Pint}).

For $n>2$ the model described by Eq.~(\ref{Zloop})
no longer displays critical points resembling branch 1 or 2,
but there exists a line of critical points \cite{ngt2tr} resembling
the hard-hexagon transition \cite{Baxhh}.

\subsection{Loop model on the square lattice}
\label{sqlat}
\begin{figure}
\includegraphics[scale=0.60]{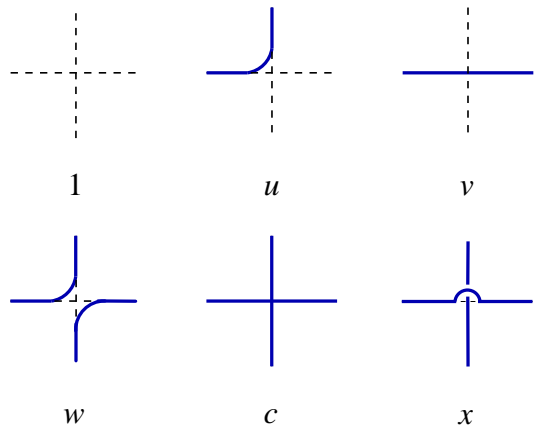}
\centering
\centering
\caption{The vertex weights $u$, $v$ and $w$ of the O($n$) loop model on
the square lattice. They are normalized such that the empty vertex has
weight 1.  In the models described by branches 1-4, the same weights
apply to rotated versions of the vertices shown here.  Also shown are
additional vertices associated with perturbations of the model of
Eq.~(\protect{\ref{Zsql}}), namely cubic vertices with weight $c$ and
crossing bond vertices with weight $x$.}
\label{sqvw}
\end{figure}

Analogous to the case of the honeycomb lattice, an O($n$) spin model
can be defined such that it can be transformed into a system of
nonintersecting loops \cite{BN,BNW} on the square lattice. In this case,
the spins are located on the middle of the edges connecting the vertices
of the O($n$) loop model. The partition function of the latter model is a 
function of the loop weight $n$ and the vertex weights $u$, $v$ and $w$.
The vertex weights of the square lattice O($n$) model are defined in
Fig.~\ref{sqvw}.

\bigskip

The partition sum of the resulting loop model is simply written in
terms of these weights as
\begin{equation}
Z_{\rm loop}=\sum_{{\mathcal G}}
u^{N_{u}} v^{N_{v}} w^{N_{w}} n^{N_{l}} \, .
\label{Zsql}
\end{equation}
The sum is on all graphs ${\mathcal G}$ consisting of nonintersecting
loops on the square lattice, and $N_{u}$, $N_{v}$ and $N_{w}$ are the
numbers of vertices with weights $u$, $v$ and $w$ respectively.

\begin{figure}
\includegraphics[scale=0.65]{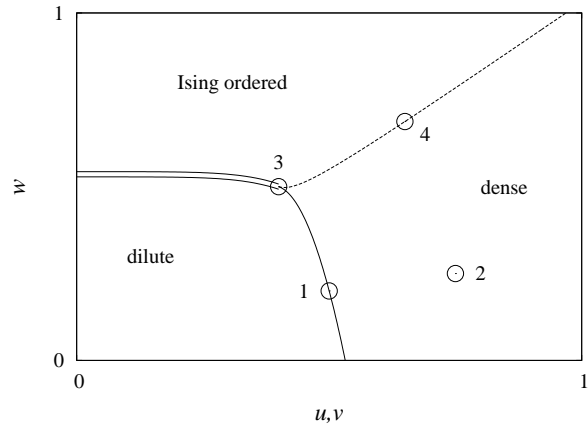}
\centering
\centering
\caption{Qualitative phase diagram of the O($n$) loop model on the square
lattice in a plane parametrized by the weights $u\approx v$ and $w$. 
The locations of the exactly solved points are indicated by the symbol
{\Large $\circ$} and the corresponding branch number.
The dilute O($n$) loop phase corresponds with the disordered phase of the
O($n$) spin model, and the dense phase with the low-temperature phase of
the spin model. The full curve represents the critical line of the O($n$)
ordering transition. The dense phase and the even denser Ising-ordered 
phase are separated by a line of Ising-like critical points, which is
shown as a dashed line. The first-order transition between the
disordered phase and the Ising-ordered low-temperature phase is shown
as a double line.}
\label{Phd}
\end{figure}

The resulting square-lattice O($n$) loop model is solvable for special
choices of the vertex weights \cite{BNW}. The solution includes four
branches of critical points, where ``critical'' refers to algebraic
decay of correlations.
These four branches form a one-parameter family, parametrized by an
angle $\theta$ that is a four-valued function of the loop weight $n$. 
For branch $k$ (with $1\leq k \leq 4$) the relation is
\begin{equation}
n= -2 \cos(2 \theta)\, \mbox{\hspace{3mm}{\rm with}\hspace{3mm}}
\frac{(2-k') \pi}{2} \leq \theta \leq \frac{(3-k') \pi}{2} \, ,
\end{equation}
with 
\begin{equation}
k'=2,~1,~3,~4  \mbox{\hspace{3mm}{\rm for}\hspace{3mm}}
k =1,~2,~3,~4  \mbox{\hspace{3mm}{\rm respectively.}}
\end{equation}
The vertex weights are
\begin{eqnarray}
u &=&\pm \, \frac{4 \sin(\theta/2) \cos(\pi/4-\theta/4)}
   {2-\{1-2\sin(\theta/2)\}\{1+2\sin(\theta/2)\}^2} \nonumber \\
v &=&\pm \, \frac{1+2\sin(\theta/2)}
   {2-\{1-2\sin(\theta/2)\}\{1+2\sin(\theta/2)\}^2} \label{sqvw0}\\
w &=&~~~\frac{1}{2-\{1-2\sin(\theta/2)\}\{1+2\sin(\theta/2)\}^2}
   \nonumber \, .
\end{eqnarray}
It appeared that, after the relabeling of $k'$ by $k$, branches 1 and 2
share the universal properties of branches 1 and 2 respectively on the
honeycomb lattice. Branch 3 represents a multicritical point where the
O($n$) critical transition,
a first-order transition, and an Ising transition merge. The Ising
degrees of freedom can be understood in terms of dual spins on the
faces of the square lattice, such that neighboring dual spins have
the same sign only if they are separated by a loop segment \cite{BN}.
The universal properties of branch 4 indicate a superposition of an
Ising-like critical state and the low-temperature O($n$) phase \cite{BN}.
Thus branch 4 is interpreted as a point where the aforementioned Ising
degrees of freedom undergo an ordering transition. A sketch of the
resulting phase diagram, as conjectured in Ref.~\onlinecite{BN} and 
confirmed in Ref.~\onlinecite{GBN} for the case $n=0$, is reproduced
in Fig.~\ref{Phd}.
\bigskip

The introduction of a sufficiently strong attractive potential between
loop segments associated with the weight $w$, can, in principle, lead to
an O($n$) tricritical point \cite{BBN}. The latter result applies only to
the case $n=0$. This tricritical point is however of a different universal
type as the multicritical point in the phase diagram  of Fig.~\ref{Phd}. 
Branches of tricritical points, parametrized by $n$, have been found for
the square \cite{GNB,NGB} and the honeycomb \cite{GBL} lattices with
vacancies.

\subsection{ The $n$-component cubic model }
\label{cubmod}
In the face-cubic spin models, the spin vector is restricted to lie
along one of $n$ Cartesian axes. Since it can still point in both
directions of each axis, it has $2n$ possible states. The spins lie
on a lattice and have nearest-neighbor couplings of the form
\begin{equation}
{\mathcal H} = - \sum_{\langle{ij}\rangle}
[K \vec{s_i} \cdot \vec{s_j} + M  (\vec{s_i} \cdot \vec{s_j})^2]\,.
\label{Hcub}
\end{equation}
A graph expansion of the $n$-component cubic model was described in
Refs.~\onlinecite{BNcub} and \onlinecite{GQBW}. The resulting partition 
sum then depends, just as in the case of the O($n$) model, continuously
on $n$. The form of the interaction between the cubic spins is, as in
Ref.~\onlinecite{GQBW}, chosen such that $e^M \cosh K=1$, in which case
the graph expansion contains only even vertices, i.e., vertices that 
connect to an even number of neighbor sites. Thus, the graph expansion 
of the cubic model introduces a new vertex with four connected legs in
comparison with the nonintersecting O($n$)  loop model.
The graph representation of the partition function of this cubic model is
\begin{equation}
Z_{{\rm cub}}=(2n)^{N} \sum_{{\mathcal G}_c} z_{\rm cub}^{N_b} n^{N_l}\,,
\label{Zcub}
\end{equation}
where $z_{\rm cub} \equiv n^{-1} e^M \sinh K $, and $N$ is the number of 
sites of the lattice, and the sum is over all graphs ${\mathcal G}_c$ that
contain only even vertices, i.e., vertices connecting to 0, 2, or 4
loop segments. The partition sum can be written in a form similar to
Eq.~(\ref{Zsql}), with $u=v=z_{{\rm cub}}$, $w=0$, and an additional
cubic four-leg vertex with weight $c=z_{\rm cub}^2$.

Also the Coulomb gas analysis of the O($n$) model \cite{CG} uses 
four-leg vertices to describe a cubic perturbation. It predicts that 
cubic perturbations are irrelevant on the O($n$) critical line for $n<2$.
Numerical analyses of the scaling dimensions of the $n$-component cubic
critical line \cite{BNcub,GQBW} do indeed confirm O($n$) universal
behavior.  In contrast, the cubic perturbation is predicted to be
relevant in the low-temperature phase.
It may thus seem rather curious that the cubic model of Eq.~(\ref{Zcub}),
when defined on the honeycomb lattice, reduces exactly to the form of
Eq.~(\ref{Zloop}), so that the cubic perturbation plays no role.
The low-temperature phase of the $n$-component cubic model of
Eq.~(\ref{Zcub}) on the honeycomb lattice is still in the universality
class of the branch-2 O($n$) model for $n<2$. The exact results for the
honeycomb O($n$) model, including the critical point, apply as well to
the $n$-component honeycomb cubic model of Eq.~(\ref{Zcub}). For the
square-lattice cubic model the critical point is not exactly known.

\subsection{ The O($n$) model with crossing bonds}
The perturbation of the low-temperature O($n$) phase by the introduction
of a square-lattice vertex with crossing bonds is predicted to be
relevant \cite{N,CG} and is thus expected to introduce crossover to 
different universal behavior. According to Jacobsen et al.~\cite{jrs},
the generic O($n$) low-temperature phase is described by the crossing-bond
model of Ref.~\onlinecite{MNR,dGN}.
This equivalence indicates the existence of a magnetic dimension $X_h=0$,
but attempts to verify this by finite-size scaling \cite{MNR,jrs} suffer
from poor convergence, which may be attributed to logarithmic factors.

\subsection{Exact exponents}
The exact results for the critical exponents of the critical and the
dense phase of the O($n$) model can be conveniently expressed in terms of
the Coulomb gas scaling dimensions $X(e_1,e_2,m_1,m_2)$ associated with
two pairs $(e_1,m_1)$ and $(e_2,m_2)$ of electric and magnetic charges.
The scaling dimension associated with these two pairs is \cite{CG}
\begin{equation}
X(e_1,e_2,m_1,m_2) = - \frac{e_1 e_2}{2 g} - \frac{m_1 m_2 g}{2}\,,
\label{Xem}
\end{equation}
where $g$ is the coupling constant of the Coulomb gas. For branches 1
and 2 of the O($n$) model it is related to the loop weight $n$ as
\begin{equation}
g= 1 \pm \frac{1}{\pi} {\rm arccos} \frac{n}{2}\,,
\label{g}
\end{equation}
where the plus sign applies to branch 1 and the minus sign to branch 2.

For the magnetic dimension $X_h$, one has $m_1=-m_2=1/2$ and
$e_1=e_2=1-g$, so that 
\begin{equation}
X_h=1-\frac{1}{2g}-\frac{3g}{8} \, .
\label{xh}
\end{equation}
In contrast with the $w$-type vertex, which is believed to be irrelevant
in the dense phase of the O($n$) loop model \cite{BN}, the cubic and the
crossing-bond vertices change the topology of the graph representation.
In the language of the mapping on the Coulomb gas, 
they are described by the four-leg watermelon diagram,
which translates  into magnetic charges $m_1=-m_2=2$ \cite{DS}.
The scaling dimensions $X_c$ of a cubic perturbation of the O($n$)
symmetry, and $X_x$ of crossing bonds, are thus
\begin{equation}
X_c=X_x=1-\frac{1}{2g}+\frac{3g}{2} \, .
\label{xc}
\end{equation}
For $n=2$ these perturbations are marginal, and for $n<2$ they are relevant
on branch 2.

The temperature dimension is \cite{CG}
\be
X_t=\frac{4}{g}-2 \, , 
\label{xtcg}
\ee
which is irrelevant for $n<2$ on branch 2. It is expected to describe the
effects of a variation of the vertex weights $u$, $v$ and $w$ with respect
to the branch-2 point.

\section{Transfer-matrix method}
\label{trmat}
The transfer-matrix technique is used to calculate the free energy
density and the magnetic correlation length of O($n$) models wrapped
on a cylinder of a finite circumference of $L$ lattice units and
of an infinite length. The free energy density of the system is
\begin{equation}
 f(L) = \frac{\zeta}{L} \ln \Lambda_0(L) \, ,
\label{flambda}
\end{equation}
where $\Lambda_0(L)$ is the leading eigenvalue of the transfer matrix
$\mat{T}$, and $\zeta$ is the geometric factor, defined as the ratio of
the unit of the finite size $L$ over the layer thickness corresponding to
the action of $\mat{T}$. Thus, for the square lattice $\zeta=1$.

It is useful to divide the transfer matrix into two diagonal blocks or
``sectors'' as follows. When one cuts the cylinder through $L$ edges
which are parallel to the axis of the cylinder, the number of dangling
loop segments may be even or odd. It is obvious that the properties of
evenness and oddness are conserved along the cylinder, so that the
transfer matrix decomposes into an odd and an even sector. The eigenvalue
$\Lambda_0(L)$ is the largest one in the even sector. The largest
eigenvalue in the odd sector is denoted $\Lambda_1(L)$. The states
of the odd sector describe the effect of an additional single loop
segment running in the length direction of the cylinder. The mapping
between the O($n$) spin model and the loop model provides the
interpretation that the odd sector describes the spin-spin correlation
function along the cylinder.
The magnetic correlation length $\xi_h(L)$ is thus
inversely proportional to the logarithm of the gap in the
eigenvalue spectrum of $\mat{T}$:
\begin{equation}
 \xi_h^{-1}(L) = \zeta \ln[\Lambda_0(L) / \Lambda_1(L)] \, .
\label{xilambda}
\end{equation}
The calculation of the eigenvalues $\Lambda_0(L)$ and $\Lambda_1(L)$ 
of $\mat{T}$ is still made subject to the condition that the associated 
eigenvectors possess translational symmetry, i.e., the eigenvectors are
invariant under the rotation of the cylinder over an angle $2\pi/L$
about its axis. The translational symmetry is in line with the form of
the partition sums given above and the periodic boundary conditions of
a model on a cylinder.
The correlation length $\xi_h(L)$ can be calculated numerically by the
transfer-matrix method as a function of a parameter $P$ representing the
distance (in some direction that remains to be specified) to a critical
point or fixed point. Including this parameter
in our notation, we define the scaled magnetic gap as
\begin{equation}
X_h(P,L)= \frac{L}{2\pi \xi_h(P,L)}\, .
\label{xlambda}
\end{equation}
For models attracted by a conformally invariant fixed point, the scaled
gap converges to the magnetic scaling dimension \cite{JCxi}. At a
distance $P$ of the fixed point, finite-size scaling \cite{FSS} then
predicts
\begin{equation}
X_h(P,L)=X_h+ aP L^{2-X_P} + \cdots\, ,
\label{xscal}
\end{equation}
where $X_p$ is the smallest scaling dimension of the scaling
fields to which $P$ contributes, $a$ is an unknown amplitude, and the
dots stand for corrections to the leading scaling behavior that vanish
for $L \to \infty$.
Since differentiation of Eq.~(\ref{xscal}) yields
\begin{equation}
\frac{d X_h(P,L)}{d P}= a L^{2-X_P} + \cdots \, ,
\label{xdiff}
\end{equation}
it is possible to estimate $X_P$ if numerical data for $d X_h(P,L)/d P$
are available for a range of finite sizes $L$. These data can be obtained
by numerical differentiation, i.e., calculation of the scaled gap for 
several values of $P$, and subsequent fitting of a polynomial in $P$
through the scaled gaps. 

The transfer-matrix construction for the O($n$) model on the square
lattice is described in Ref.~\onlinecite{BN}, including the coding that
defines the transfer-matrix index in terms of the ``connectivities''
describing the topology of the loop configuration at a cross-section of
the cylinder. A sparse-matrix decomposition allows the evaluation of
the leading eigenvalues of transfer matrices with linear sizes up to a
few times $10^7$ with the use of modest computer resources. Calculations
for the O($n$) model on the honeycomb lattice require a different
sparse-matrix decomposition, which is explained in Ref.~\onlinecite{BNP}.

Furthermore, we shall also introduce two new types of vertices on the
square lattice that generate a larger set of connectivities than those
of the nonintersecting loop model.
These are the cubic vertex and the crossing-bond vertex, included in
Fig.~\ref{sqvw}.
The introduction of the cubic vertex into the O($n$) loop model leads
to connections between the loops and thus leads to a larger set of
connectivities in comparison with the loop model. The number of mutually
connected dangling edges is no longer restricted to two but may also
assume multiples of 2.
The coding and decoding needed for the construction of the transfer
matrix corresponding with the cubic model of Eq.~(\ref{Zcub}) was
described in Ref.~\onlinecite{BNcub}.

Also the presence of crossing-bonds leads to an increase of the number
of connectivities. While the dangling loop segments can only be connected
pairwise, the ``well-nestedness'' property of the
non-intersecting loop model is lost. This property implies that,
if dangling bonds $i$ and $j$ are connected, and dangling bonds $k$ and
$l$ are also connected, that the situation $i<k<j<l$ is excluded. Once
crossing-bond vertices are allowed, the loops get entangled, and the
situation $i<k<j<l$ becomes possible.
The coding of this larger set of connectivities by means of integers 1,2,
$\cdots$ is actually simpler than that of the well-nested connectivities
\cite{BN}. The coding is determined by a set of rules specifying an
ordering of these connectivities. For completeness, we describe an ordering
including the non-well-nested $L$-point connectivities. It is useful to
represent a connectivity $\alpha$ by an array of integers 
$\vec{i}_{\alpha} \equiv (i_{\alpha}(1),i_{\alpha}(2),\ldots,i_{\alpha}(L))$,
such that $i_{\alpha}(k)=i_{\alpha}(l)$
if and only if the positions $k$ and $l$ are connected, i.e.,
if dangling edges $k$ and $l$ are covered by dangling segments of
the same loop. The special value $i_m=0$ represents a dangling edge not
covered by a loop segment.
The ordering of the connectivities, denoted by Greek symbols, is
formulated in terms of these arrays of integers.  The rules are:
\begin{enumerate}
\item
For the $L_\alpha$-point connectivity $\alpha$, remove the integers
with $i_m=0$ from the array $\vec{i}_{\alpha}$. This leads to an
$L_{\tilde{\alpha}}$-point dense connectivity $L_{\tilde{\alpha}}$
without vacancies, represented by an array $\vec{i}_{\tilde{\alpha}}$.
Then, connectivity $\alpha$ precedes connectivity $\beta$ if
$L_{\tilde{\alpha}}$ exceeds the corresponding number $L_{\tilde{\beta}}$
of connectivity
$\beta$.  This provides only a partial ordering; it remains to order
the set of connectivities with $L_{\tilde{\alpha}}=L_{\tilde{\beta}}$.
This remaining ordering will depend on the positions
of the zeroes in $\vec{i}_{\alpha}$, and on 
the connectivity $\tilde{\alpha}$ of the remaining dense configuration.
\item 
Form an $L_{\alpha}$-bit binary number $B_{\alpha}$ with 0 (1) on 
position $k$ if $i_{\alpha}(k) \neq 0$ ($i_{\alpha}(k) = 0)$. We can now
specify that connectivity $\alpha$ precedes connectivity $\beta$
if $L_\alpha=L_\beta$ and $B_{\alpha}<B_{\beta}$. 
\item The remaining task is to order the dense connectivities
$\tilde{\alpha}$. The first part is to find the position
$n_{\tilde{\alpha}}$ of the loop segment connecting to the loop segment
on position 1, i.e., the number that satisfies
$i_{\tilde{\alpha}}(1) = i_{\tilde{\alpha}}(n_{\tilde{\alpha}})$.
Then, we specify that dense connectivity $\tilde{\alpha}$ precedes 
$\tilde{\beta}$ if $n_{\tilde{\alpha}}<n_{\tilde{\beta}}$.
\item  If $n_{\tilde{\alpha}}=n_{\tilde{\beta}}$, we define an 
$L_{\tilde{\alpha}'} =L_{\tilde{\alpha}}-2$-point dense connectivity
$\tilde{\alpha}'$ by removing positions 1 and $n_{\tilde{\alpha}}$ from
$\tilde{\alpha}$. The remaining ordering is provided by the recursive
application of the last two steps, adding primes at each new iteration,
until a decision is found. 
\end{enumerate}
The enumeration on the basis of this ordering requires only some trivial
bookkeeping, involving numbers of connectivities of the relevant types,
using methods presented already in Refs.~\onlinecite{BN} and
\onlinecite{BN82}. An inverse algorithm that derives an array
$\vec{i}_{\alpha}$ for a given connectivity number or transfer matrix
index, was constructed similarly.

\section{Crossover and the dense phase}
\label{numres}
We investigate the influence of various perturbations with respect
to the branch-2 models on the honeycomb and the square lattice.

\subsection{Attractions between loop segments on the square lattice}    
\label{paor}
We choose the plus signs in Eq.~(\ref{sqvw0}), while noting that the
sign of $u$ is irrelevant because the number of $u$-type vertices is
even in the systems of interest. Since the number of $v$-type vertices
in a loop wrapping a cylinder with odd $L$ is also odd, we have to 
keep in mind that the sign of $v$ matters for odd system sizes. 

Denoting the vertex weights at the branch-2 and the branch-4 points as
$(u^{(2)},v^{(2)},w^{(2)})$ and $(u^{(4)},v^{(4)},w^{(4)})$ respectively,
we interpolate between the branch-2 and branch-4 points, and also
extrapolate, by varying $p$ in
\begin{eqnarray}
u(p)  &=&  (1-p) u^{(2)} + p u^{(4)}     \nonumber \\
v(p)  &=&  (1-p) v^{(2)} + p v^{(4)}     \label{clt}\\
w(p)  &=&  (1-p) w^{(2)} + p w^{(4)}\, . \nonumber
\end{eqnarray}

\begin{figure}
\includegraphics[scale=0.37]{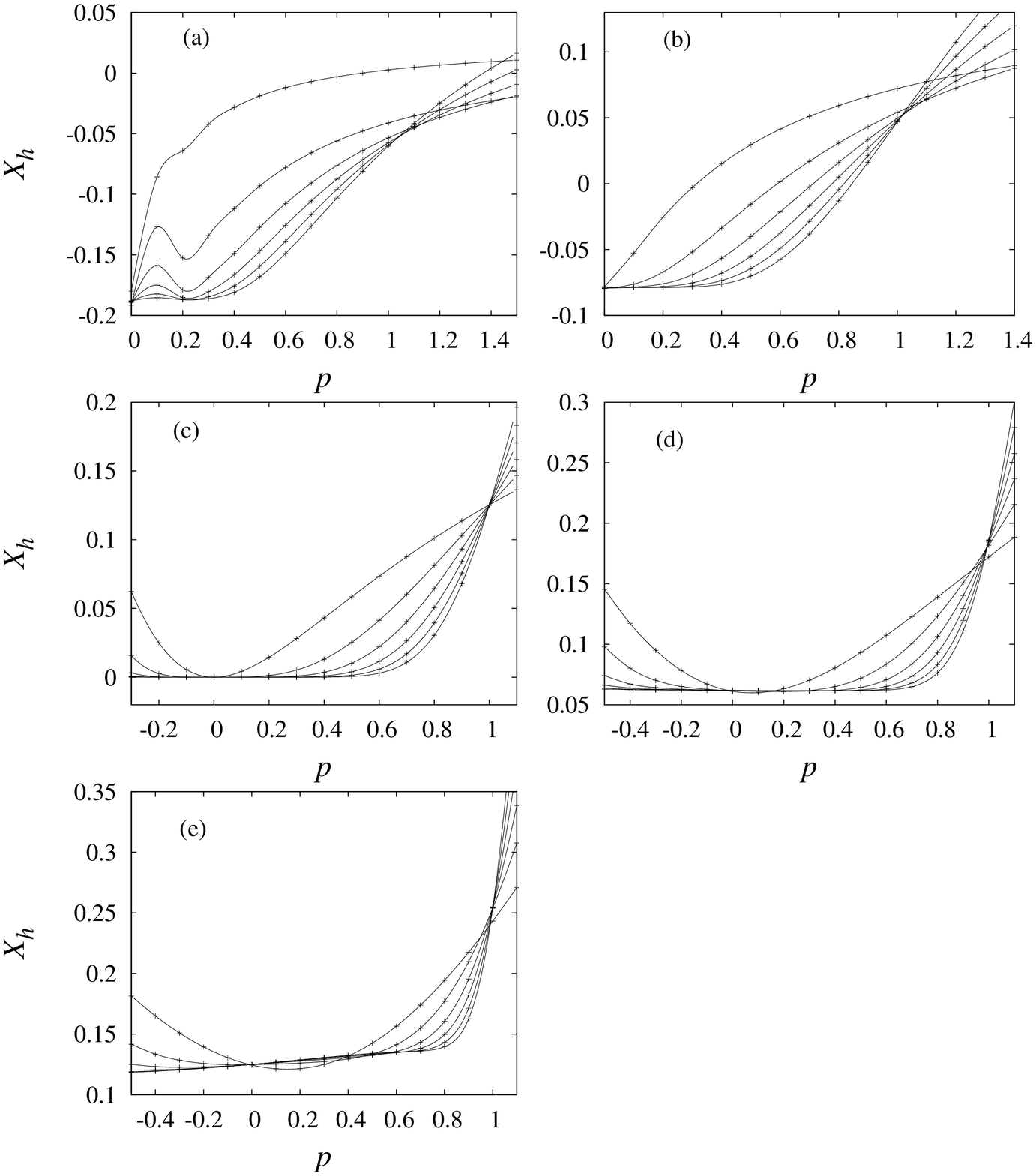} 
\caption{ Scaled magnetic gap $X_h$ versus the parameter $p$
that interpolates between branch 2 ($p=0$) and branch 4 ($p=1$) for
the O($n$) model. (a) $n=0$, (b) $n=0.5$, (c) $n=1$, (d) $n=1.5$, and (e)
$n=2$.  Results are shown for finite
sizes $L=2$, 4, $\cdots$, 12. The slope of the curves 
increases with $L$ on the right-hand side.}
\label{int24n00}
\end{figure}

We varied $p$ in the range $-0.5\leq p \leq 1.5$ and calculated the
scaled magnetic gap $X_h(p,L)$. Plots of this quantity as a function of $p$,
for several values of the finite-size parameter $L$, are shown in 
Fig.~\ref{int24n00}. 
The main effect of increasing $p$ is that the weight $w$, which controls
the attraction between neighboring loops, also increases.

As a result, the loop configuration becomes denser, and at $p=1$ it
condenses into a state with Ising order as mentioned in Sec.~\ref{sqlat}. 
One observes that, for most $n$, there are clearly two different
intersections of the curves for different $L$, near $p=0$ and $p=1$.
For $p=0$ the steepest curves are those with the smallest $L$
of the curves for different $L$. For $p=1$ this situation is just the
reverse. 
This shows that the perturbation with respect to branch 2 is irrelevant,
in agreement with the expected behavior for the leading thermal
exponent according to Eq.~(\ref{xtcg}).
In contrast, the perturbation due to the variation of $p$ with respect
to the branch-4 point is seen to be relevant. Numerical differentiation
to $p$ of the scaled gaps at the branch-4 points for several $n$, and
subsequent analysis according to Eq.~(\ref{xdiff}) (with $p$ instead of
$P$) yielded estimates of $X_p$ that are shown in Table \ref{numdx}. 
One observes that the results for $X_p$ are close to the known Ising 
temperature dimension $X_t=1$. This confirms the Ising nature of the
transition driven by $p$, in line with the conclusion \cite{BN} that
it takes place independent of the critical background of the dense
O($n$) model.

\begin{table}
\caption{Numerical estimates of the scaling dimensions $X_p$ and
$X_{w_2}$, which belong to the most relevant scaling fields to which
$p$ and $w_2$ contribute respectively.
}
\begin{tabular}{|c|c|c|}
\hline
$n$   & $X_p$       & $X_{w_2}$\\
\hline
$0  $ & 1.000~(1)   & 1.0(1)\\
$0.5$ & 1.0000~(1)  & 0.99(1)\\
$1  $ & 1.000000~(2)& 0.9999(1)\\
$1.5$ & 1.00001~(1) & 1.001(1)\\
$2$   & 1.01~(1)    & 0.999(1)\\
\hline
\end{tabular}
\label{numdx}
\end{table}

\subsection{The nature of the Ising-ordered phase}
\label{Xm}
Figure \ref{int24n00} 
in the preceding subsection, and 
numerical results for larger values of $p>1$, indicate that the scaled 
gaps increase approximately linearly with the finite size, and thus
that the magnetic correlation length becomes constant.
This corresponds with a magnetic correlation function that decays
exponentially in the infinite plane. In this respect, the dense phase
and the Ising ordered phase, separated by the Ising line as shown in
Fig.~\ref{Phd}, are different. Still, these two phases
are assumed \cite{BN} to share the basic universal properties of the
dense O($n$) loop model. 
We test this assumption by eliminating the reason why the scaled gaps
increase sharply at $p=1$ and beyond. The reason is that, near the line
of Ising transitions, the loop configurations become so dense that most
vertices are of the $w$-type. Since the even sector of the transfer
matrix allows only loop configurations that cover an even number of edges
in the transfer direction, only configurations of the even sector fit
well on a lattice with even $L$. This explains the increase of the gap
between the even and the odd sector as the the Ising line is approached.
Similarly, such dense configurations in the odd sector will only fit
well on lattices with odd $L$.

In order to define a type of magnetic gap that excludes these effects
of even-odd alternation, one has to select even or odd systems in
accordance with the sector. We thus define a magnetic scaled gap
$X_m(p,L)$ as 
\be
X_m(p,L)\equiv \frac{\zeta L}{2\pi} 
\ln \frac{\sqrt{\Lambda_0(L+1) \Lambda_0(L-1)}}{\Lambda_1(L)}
\ee
for odd $L$, and
\be
X_m(p,L)\equiv \frac{\zeta L}{2\pi} 
\ln \frac{\Lambda_0(L)}{\sqrt{\Lambda_1(L+1) \Lambda_1(L-1)}}
\ee
for even $L$.

\begin{figure}
\includegraphics[scale=0.37]{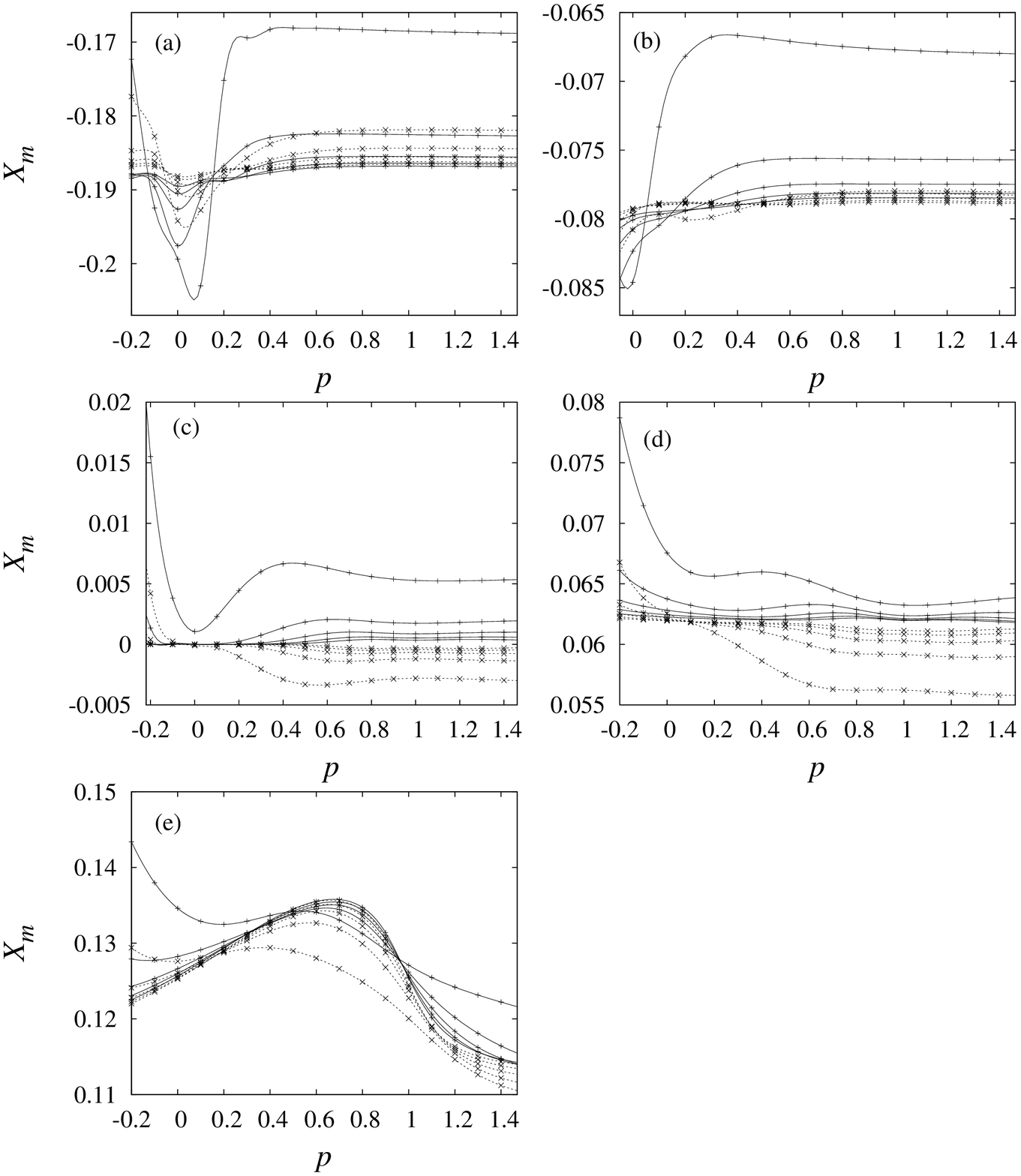}
\caption{ Scaled magnetic gap $X_m(p,L)$ versus the 
parameter $p$ that interpolates between branch 2 ($p=0$) and branch 4
($p=1$) for the cases (a) $n=0$, (b) $n=0.5$, (c) $n=1$, (d) $n=1.5$, 
and (e) $n=2$  O($n$) model. Results are shown for $L=3$, 4,
$\cdots$, 12. The dashed curves with symbols $\times$ are used for even
$L$, and the solid curves with symbols $+$ for odd $L$. Both sets of curves
tend to converge to a common $p$-independent limit when $L$ increases.}
\label{xm24n00}
\end{figure}

The results for these scaled magnetic gaps are shown in Fig.~\ref{xm24n00} 
as a function of $p$, for several values of $n$.
These results indicate that, for $n<2$, the universal character
of the magnetic correlations in the dense phase is independent of $p$.
In particular, it remains unchanged under
the Ising transition and the onset of the Ising-type long-range order.

Next we estimated the associated magnetic scaling dimension $X_m$ at 
the Ising point (branch 4) by fitting the numerical results using
Eq.~(\ref{xscal}) for several $n$.  The results are listed in
Table \ref{xh1}. Here our choice of the sign of $v$ in Eq.~(\ref{sqvw0}) 
follows the change of sign of $w$ such that $v w>0$.
The change of sign of $w$ is caused by the change of sign of the
common denominator in Eq.~(\ref{sqvw0}), which arises because the weight 
of the empty vertex is normalized to +1.
A consequence of this change of sign is that there will be a jump with
value 1/8 in $X_m$ near $n=0.087378025$, where the change of sign of $w$ 
occurs, if the weight $v$ is kept positive. The jump is equal to the
``interface dimension'' denoted $X_{{\rm int},1}$ in Ref.~\onlinecite{BN}.

\begin{table}
\caption{Numerical results for the magnetic scaling dimension $X_m$ of
branch 4 (rightmost column), compared to the exact results for the
magnetic dimension $X_h$ for branch 2 and for branch 4. The branch-4
magnetic dimension is equal to the branch-2 dimension plus $1/8$.
}
\label{xh1}
\begin{tabular}{|c|c|c||c|}
\hline
$n$     &$X_h$ (branch 2)&$X_h$ (branch 4)  & $X_m$           \\
\hline
$0$     & $-0.1875    $ & $-0.0625       $  &$-0.18749~(1)  $ \\
$0.075$ & $-0.16865534$ & $-0.04365534312$  &$-0.168653~(2) $ \\
$0.0875$& $-0.16561812$ & $-0.04061812640$  &$-0.16562~(2)  $ \\
$0.1  $ & $-0.16260929$ & $-0.03760929546$  &$-0.162610~(2) $ \\
$0.125$ & $-0.15667507$ & $-0.03167507924$  &$-0.156675~(2) $ \\
$0.5$   & $-0.07909087$ & $ 0.04590912236$  &$-0.07909~(1)  $ \\
$1  $   & $0   $        & $ 0.125        $  &$ 0.0000000~(1)$ \\ 
$1.5$   & $0.061874313$ & $ 0.18687431332$  &$ 0.0618743~(1)$ \\
$2$     & $0.125$       & $ 0.25         $  &$ 0.125000~(1) $ \\
\hline
\end{tabular}
\end{table}

\subsection{Double bonds in the honeycomb model}
\label{dbhc}
\begin{figure}
\includegraphics[scale=0.16]{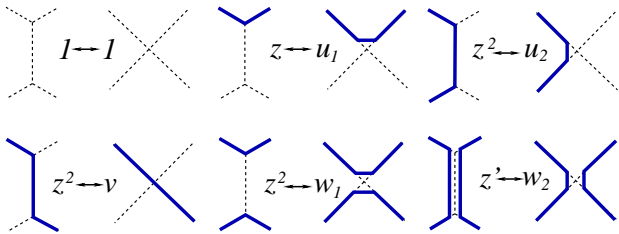}
\centering
\centering
\caption{The vertex weights of the O($n$) loop model on the square
lattice as obtained by a mapping on the honeycomb lattice. These
vertices allow the occupation of vertical edges of the honeycomb
lattice by two loop segments. The resulting vertex weights allow for
a factor $z$ per loop segment on the honeycomb lattice.}
\label{mapsqhc}
\end{figure}

The loop model of Eq.~(\ref{Zloop}) is extended by allowing the edges
with one of the three possible orientations, say the vertical edges,
to be covered by up to two loop segments. The honeycomb lattice can be
decomposed in building blocks consisting of a vertical edge and the pair
of vertices at its ends. These units are shown in Fig.~\ref{mapsqhc},
together with their weights, which include a factor $z$ per loop segment.

\bigskip
As indicated in Fig.~\ref{mapsqhc}, each such unit of the honeycomb
lattice can be replaced by a square-lattice vertex. This substitution
maps the honeycomb model with double bonds on an O($n$) loop model on the
square lattice, but the vertex weights, also shown in Fig.~\ref{mapsqhc},
are not of the form of Eq.~(\ref{Zsql}) because the weights of the $u$-
and $w$-type vertices depend on their orientation.  Expressed in the
enlarged set of square-lattice weights, the partition sum takes the form
\begin{equation} Z_{\rm loop}=\sum_{{\mathcal G}} u_1^{N_{u_1}}
u_2^{N_{u_2}} v^{N_{v}} w_1^{N_{w_1}} w_2^{N_{w_2}} n^{N_{l}} \, ,
\label{Zsq2}
\end{equation}
where the indices appended to $u$ and $w$ indicate the orientation
of the vertex.
The vertex weight $z'=w_2$ describes a double bond covering a lattice
edge. The resulting connectivities are, however, still of the
nonintersecting loop type, which means that the two loop segments on
an edge do not cross or mutually connect.

\begin{figure}
\includegraphics[scale=0.37]{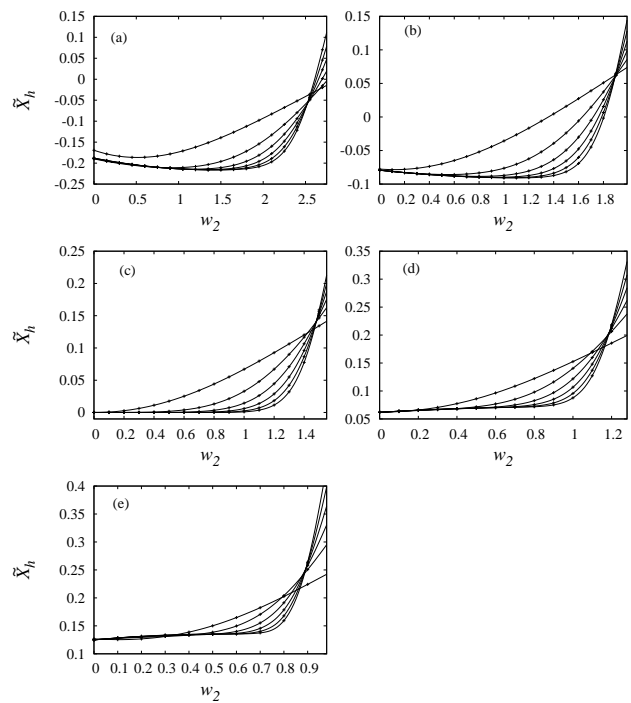}
\caption{ Scaled magnetic gap $\tilde{X}_h(w_2,L)$ versus 
the weight $w_2$ of a double bond covering an edge of the honeycomb
lattice for O($n$) model with (a) $n=0$, (b) $n=0.5$, (c) $n=1$, 
(d) $n=1.5$, (e) $n=2$. 
Results are shown for finite
sizes $L=2$, 4, $\cdots$, 12. The slope of the curves
increases with $L$ on the right-hand side.}
\label{dbhc00}
\end{figure}

We evaluated the scaled gaps according to  Eqs.~(\ref{xilambda}) and
(\ref{xlambda}) of the model of  Eq.~(\ref{Zsq2}) for several values of
the finite size $L$, using the weights $u_1$, $u_2$, $v$ and $w_1$ as
obtained from the equivalence with the branch-2 point of the honeycomb
model with $z$ according to Eq.~(\ref{hczc}). The influence of double
bond bonds was determined by including several nonzero values of the 
weight $z'=w_2$.
The calculation of the scaled gaps used the geometric factor
$\zeta=2/\sqrt{3}$ for the honeycomb lattice. Since the additional
weight $z'$ introduces anisotropy, the asymptotic conformal symmetry 
is broken, and the scaled gaps for $z'\neq 0$ no longer directly relate
to the scaling dimension $X_h$. For this reason we add a tilde and
denote the scaled gaps as $\tilde{X}_h(z',L)$.
The results are shown in Fig.~\ref{dbhc00}. 

While the interpretation of the scaled gap  in terms of the scaling
dimension $X_h$ is no longer valid, the intersections on the right-hand
side of these figures, with slopes increasing with $L$, still indicate
that a phase transition takes place, resembling the Ising-like ordering
for the square lattice model in Sec.~\ref{paor}.
The Ising character of this transition was verified by means of
numerical differentiation of $\tilde{X}_h(w_2,L)$ with respect to
$w_2$ in the intersection points and finite-size scaling, analogous
to the analysis of $X_p$ in Table \ref{numdx}. The numerical estimates
of the scaling dimension $X_{w_2}$ for several $n$ are included in 
Table \ref{numdx}. They are close to the Ising temperature 
dimension $X_t=1$.

\subsection{Cubic anisotropy}
\label{cubani}
The introduction of cubic vertices into the O($n$) model of 
Eq.~(\ref{Zsql}) modifies the partition function as follows
\begin{equation}
Z_{\rm loop}=\sum_{{\mathcal G}}
u^{N_{u}} v^{N_{v}} w^{N_{w}} c^{N_{c}} n^{N_{l}} \, ,
\label{Zlcu}
\end{equation}
where $N_{c}$ is the number of vertices of type $c$. For $w=0$, $v=u$ and
$c=u^2$ it reduces, apart from a multiplicative constant, to the partition
sum of the cubic model, Eq.~(\ref{Zcub}). Thus, Eq.~(\ref{Zlcu}) can
interpolate between the nonintersecting loop model and the cubic model.

\begin{figure}
\includegraphics[scale=0.37]{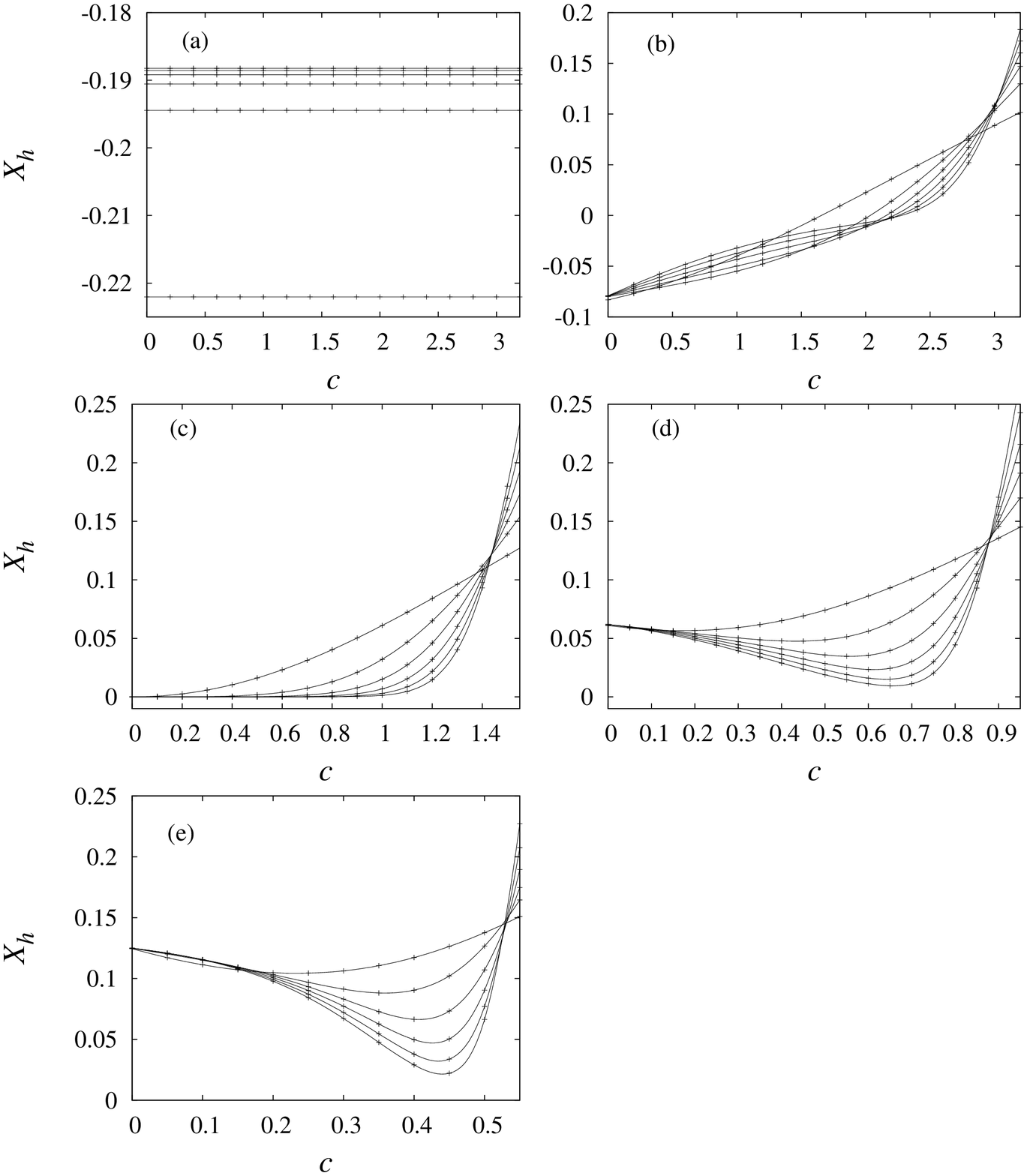}
\caption{Scaled magnetic gap $X_h$ versus the weight
$c$ of cubic vertices introduced into the dense phase of the O($n$)
nonintersecting loop model according to Eq.~(\protect{\ref{Zlcu}}).
Results are shown for finite sizes $L=2$, 4, $\cdots$, 12. (a) $n=0$.
The weight $c$ has no effect because the loop weight $n=0$ does not
allow a nonzero density of $c$-type vertices. 
The gap increases as a function of $L$. The cases $n=0.5,1,1.5,2$
are shown in (b)-(e) respectively.
The slope of the curves increases with $L$ on the right-hand side.}
\label{cver00}
\end{figure}

We investigate the effect of cubic perturbations by varying $c$,
while keeping the other vertex weights fixed at their branch-2 values.
The results for the scaled gaps are shown in Fig.~\ref{cver00}. 
For $n=0$, the vanishing loop weight prevents the
introduction of cubic vertices, so that the results do not depend on
$c$. The slopes of the curves with $0<n<2$ are seen to increase with
$L$ near $c=0$, which shows that the  cubic perturbation is relevant.
Furthermore, some of the plots display two more sets of intersections, 
of which the middle ones are indicative of a stable fixed point,
and the rightmost ones of an unstable fixed point, resembling that
of the Ising-like transition induced by the $w$-type vertices.
These results are consistent with the interpretation that,
in the range attracted by the stable fixed point, the scaled gap
converges to $X_h=0$ for $0<n<2$ in a range $c>0$. This could be
confirmed by numerical extrapolations at some values of $c$,
all of which satisfied $X_h \leq 0.01$.

In order to numerically determine the exponent responsible for the
cubic crossover, we performed a numerical differentiation of the scaled 
magnetic gap $X_h(c,L)$ with respect to the weight $c$ of the cubic
vertex at the point $c=0$ for several values of $n$.
The finite-size data for this derivative were subsequently analyzed 
according to Eq.~(\ref{xdiff}), with $P$ replaced by $c$.
The resulting estimates of $X_c$ are shown in Table \ref{Xcx},
together with the Coulomb gas predictions.

\begin{table}
\caption{
Scaling dimensions $X_c$ associated with a cubic perturbation,
and $X_x$ associated with crossing bonds for the case of the dense
O($n$) phase. These results are obtained by numerical differentiation
of the scaled magnetic gap $X_h(L)$ at the branch 2 point.
The vanishing of some numerical results for $n=0$ and 1 is an artefact
due to suppression of the respective critical amplitudes. 
}
\begin{tabular}{|c|c|c||c|c|}
\hline
\multicolumn{1}{|c|}{}&\multicolumn{2}{c||}{Coulomb gas (branch 2)}&
\multicolumn{2}{c|}{numerical results} \\
\hline
$n$   & $g$      & $X_x=X_c$ & $X_c$   &  $X_x$      \\
\hline
$0  $ & $0.5 $   & $0.75 $   &   0     & 0.75001~(1)  \\
$0.5$ & $0.58043$& $1.00922$ &1.009~(1)& 1.0092~(2)   \\
$1  $ & $2/3     $ & $1.25$  &   0     & 0            \\
$1.5$ & $0.76995$& $1.50552$ &1.505~(1)& 1.5055~(3)   \\
$2$   & $1$      & $2$       &2.000~(2)& 1.998~(2)    \\
\hline
\end{tabular}
\label{Xcx}
\end{table}

\subsection{Crossing bonds} 
\label{crosb}
\begin{figure}
\includegraphics[scale=0.37]{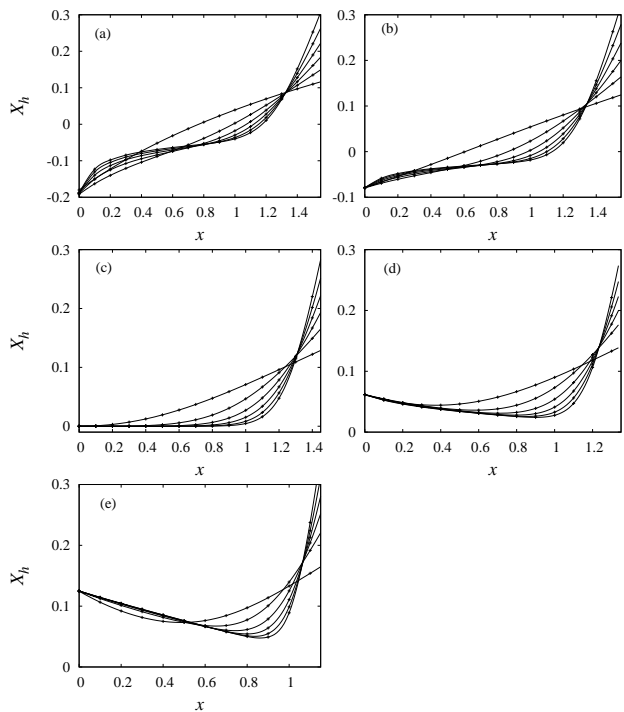}
\centering
\centering
\caption{ Scaled magnetic gap $X_h$ versus the weight of the
crossing-bond vertex, for the cases (a) $n=0$, (b) $n=0.5$, (c) $n=1$,
(d) $n=1.5$, (e) $n=2$.  Results are shown for finite
sizes $L=2$, 4, $\cdots$, 12. The slope of the curves
increases with $L$ on the right-hand side.}
\label{sqcr00}
\end{figure}

We next introduce, starting from the branch-2 low-temperature points of 
the square-lattice loop model, a nonzero weight of the crossing-bond vertex.
We calculated the scaled gaps for a range of values of the weight $x$, and
for several values of $n$.
The results are shown in Fig.~\ref{sqcr00}. 
These data indicate that, for $n<2$ and a range of $x>0$, crossover
occurs to a different universality class of dense intersecting loop
models, with a magnetic exponent that is different from that of
nonintersecting loop models for $n \neq 1$. This interpretation is in
line with a prediction of Jacobsen et al.~\cite{jrs} in terms of
exact results \cite{MNR,dGN}. We have attempted to find the conformal 
anomaly $c_a$ and the magnetic exponent $X_h$ from the finite-size
data for a few points in the phase diagram. Estimates for $c_a$ are
obtained by fitting $f(L)=f(\infty) +\pi c_a/(6 L^2)$ to transfer-matrix
results for the free energy using three subsequent values of $L$.
These estimates display slow apparent convergence and are thus hard to
extrapolate. Extrapolation was done assuming finite-size dependence
as $L^{-2}$. The results are listed in Table \ref{cxx}. As a tentative
error margin we quote ten times the difference between the last two
extrapolations. Also the data for $X_h$ were hard to extrapolate;
we simply quote the result $X_h(L)$ obtained from Eq.~(\ref{xlambda})
for $L=14$, with a tentative error margin of 10 times $X_h(14)-X_h(12)$.
\begin{table}
\caption{Numerical estimates of the conformal anomaly $c_a$ and the
magnetic exponent $X_h$ in the dense O($n$) phase with crossing bonds.
}
\begin{tabular}{|c|c|c|c|}
\hline
$n$   &  $x$  &   $c_a$        &  $X_h$           \\
\hline
$0  $ & 0.5   &  $-1.42$ ~(26) & $-0.066$ ~(29)   \\
$0  $ & 0.8   &  $-1.36$ ~(29) & $-0.053$ ~(15)   \\
$0.5$ & 0.4   &  $-0.64$ ~(19) & $-0.036$ ~(13)   \\
$0.5$ & 0.6   &  $-0.65$ ~(10) & $-0.031$ ~(~9)   \\
$1  $ & 0.8   &  $-0.01$ ~(10) &~$~0.000$ ~(~2)   \\
$1  $ & 1.1   & ~$~0.05$ ~(64) &~$~0.011$ ~(50)   \\
$1.5$ & 0.6   & ~$~0.55$ ~(~2) &~$~0.031$ ~(~3)   \\
$1.5$ & 0.8   & ~$~0.53$ ~(11) &~$~0.025$ ~(~3)   \\
$1.5$ & 1.0   & ~$~0.52$ ~(35) &~$~0.025$ ~(30)   \\
$2$   & 0.3   & ~$1.000$ ~(~2) &~$~0.095$ ~(~1)   \\
$2$   & 0.6   & ~$0.998$ ~(46) &~$~0.067$ ~(~1)   \\
$2$   & 0.9   & ~$~0.97$ ~(29) &~$~0.046$ ~(35)   \\
\hline
\end{tabular}
\label{cxx}
\end{table}

We also performed numerical differentiations of the scaled magnetic
gaps $X_h(x,L)$ with respect to the weight $x$ of the crossing-bonds
vertex at the point $x=0$ for several $n$.
The finite-size data were subsequently analyzed according to
Eq.~(\ref{xdiff}), with $P$ replaced by $x$.
This yielded estimates of $X_x$ that are shown in Table \ref{Xcx},
together with the Coulomb gas predictions.

\section{Discussion}
\label{disc}
The numerical results presented in Sec.~\ref{paor} for the effect
of the $w$-type vertex, representing loop-loop attractions, on the
low-temperature phase of the square-lattice model, agree with the
predicted \cite{BN} behavior. As indicated by the curves in 
Fig.~\ref{int24n00}, these attractions are irrelevant in
the whole range interpolating between the branch-4 and branch-2 points.
In contrast, they are relevant in the equivalent tricritical O($n$)
model with vacancies \cite{NGB}. The latter model is obtained by
summing out part of the loops, which thus yields a system that is far 
more susceptible to attractions between the loops.

Furthermore, the numerical results confirm that the loop-loop attractions
are also relevant at the branch-4 point, and that the transition that
takes place at this point is Ising-like. Apart from the
explanation of this transition in terms of the onset of long-range
order of the dual Ising spins, it may be worthwhile to mention that
the type of phase diagram in Fig.~\ref{Phd}, including the Ising line,
is reproduced by an Ising model with vacancies quoted in
Ref.~\onlinecite{NGB}. In the latter case, the Ising line
corresponds with the onset of phase separation between a phase
dominated by Ising spins and a phase dominated by vacancies.

The analysis of the O($n$) magnetic dimension $X_m$ presented in
Sec.~\ref{Xm} confirms that the low-temperature O($n$) universal 
character remains unaffected across the Ising transition on branch 4.
This result was obtained by means of a careful formulation of the
correlation function associated with $X_m$, which takes into account 
the even- or oddness of the system, even in the limit of infinite size.

On the basis of results \cite{jrs,MNR,dGN} for loop models that allow
crossing bonds, doubts have arisen to what extent the behavior found
for the exactly solvable models of branch 2 of the honeycomb \cite{N},
the square \cite{BNW}, and the triangular lattice \cite{KNB} is 
representative for the low-temperature O($n$) phase. Since multiple
bonds arise in a natural way in graph expansions of more general spin
O($n$) models, we studied the effect of double bonds in Sec.~\ref{dbhc}.
This perturbation was found to be irrelevant for branch 2 of the model
on the honeycomb lattice. Its effect appears to be very similar to that
of loop-loop attractions due to the type-$w$ vertex in the case of the
square lattice. The similarity includes the Ising-like transition that
takes place at a sufficient weight of the double bonds.

The Coulomb gas prediction that the cubic perturbation is relevant for
$n<2$ is quantitatively confirmed by the results for $X_c$ in
Table \ref{Xcx} only for $n=0.5$ and 1.5. We attribute the vanishing
numerical results for $n=0$ and 1 to vanishing amplitudes associated
with the cubic perturbation.  The numerical result $X_c=0$ at $n=0$ is
due to the fact that the zero loop weight excludes type-$c$ vertices
even at nonzero fugacity. For $n=1$, the loop weights are equal to 1,
so that the distinction between type-$z$ and type-$c$ vertices disappears,
and so do the amplitudes associated with the cubic perturbation.
Taking into account this explanation for the deviating results for
$n=0$ and 1, we conclude that there is a satisfactory agreement with
the Coulomb gas predictions.

As mentioned in Sec.~\ref{cubmod}, the relevance of cubic perturbations
in the low-temperature phase of the square-lattice model of
Eq.~(\ref{Zcub}) for $n<2$, which was confirmed in Sec.~\ref{cubani},
seems very peculiar in relation with the absence of crossover to cubic
behavior for the  honeycomb model. In view of the mutually incompatible
values of $X_h$, the conclusion that the dense phases
of the partition sum of Eq.~(\ref{Zcub}) display different universal
behavior for the square and honeycomb lattices is inescapable, and has
to be attributed to the low coordination number of the honeycomb lattice.

One may wonder if a similar paradox occurs in the spin representation of
the model, for which our physical intuition may provide further insight.
The equivalence with Eq.~(\ref{Zcub}) applies only to cases where $n$
is a positive integer. The case $n=1$ fails to provide more clarity   
because the amplitude associated with $c$ vanishes. For $n=2$ we do
expect a phase transition to a long-range-ordered state in the spin
model described by Eq.~(\ref{Hcub}) at sufficiently low temperatures,
but the condition $e^M \cosh K=1$ excludes this low-temperature range
from the loop representation of Eq.~(\ref{Zcub}). There is no apparent
conflict with the expected behavior of the cubic spin model.

The introduction of the cubic vertices into the loop model on the
square lattice in Sec.~\ref{cubani} yielded
results that are consistent with the interpretation that the scaled
gaps converge to 0 for $n<2$ in a range $c>0$.  This interpretation
is in line with the expected long-range order of spin models with a
cubic perturbation at low temperatures.

Furthermore we note that the mapping of the honeycomb model on the
square lattice model presented in Sec.~\ref{dbhc} can be extended to 
include cubic vertices on the square lattices, which corresponds to
mutually connecting double bonds on the vertical edges of the
honeycomb model. Thus, cubic crossover will occur on a suitably
generalized honeycomb model.

Also in the case of perturbations introduced by the crossing-bond vertex
we find a satisfactory agreement with the Coulomb gas predictions.
Also in this case the amplitude $a$ in Eq.~(\ref{xscal}) due to the
perturbation vanishes at $n=1$ (see Table \ref{Xcx}), which explains
the vanishing of the corresponding numerical result. For $n=1.5$ the
amplitude is still rather small (see Fig.~\ref{sqcr00}(d)), but the
numerical differentiation method is sufficiently sensitive to determine
the scaling dimension $X_x$.

The scaled gaps in the dense phase perturbed by crossing bonds appear
difficult to analyze. While our range of finite sizes is insufficient
for reliable extrapolations, the data seem in line with $X_h=0$ and the
occurrence of logarithmic factors as predicted for crossing-bond
models \cite{MNR,dGN,jrs}. Moreover, the results for the conformal
anomaly given in Table \ref{cxx} show a trend consistent with
$c_a=n-1$ as predicted by Martins et al.~\cite{MNR}.

In conclusion, our results confirm the phase diagram of the 
O($n$) model as conjectured in Ref.~\onlinecite{BN}, in particular the
Ising transition between the dense and the Ising-ordered phases,
and its Ising scaling dimension. Cubic anisotropy and crossing 
bonds are proved numerically to be relevant and introduce crossover to 
different universal behavior in the low-temperature (dense) phase.

Finally, we note that our numerical results display so-called
nonuniversal behavior as a function of the various perturbations
in the $n=2$ models, as expected from the mapping between the cubic
model and, e.g., the Ashkin-Teller model \cite{BNcub} and the equivalence
of the latter model with the eight-vertex model \cite{Fan,Weg,Bax}.

\acknowledgments
We are indebted to Prof. B. Nienhuis for valuable discussions.
W. G. acknowledges hospitality extended to him by the Lorentz Institute.
This work is supported by the Lorentz Fund, by the NSFC under Grant No.
10675021, the NCET, and by the HSCC (High Performance Scientific Computing
Center) of the Beijing Normal University.

\end{document}